\documentclass[a4paper,a4wide, 11pt]{scrartcl}
\usepackage{a4wide}
\usepackage{epsfig}
\usepackage{latexsym}
\usepackage{amssymb}
\usepackage{amsmath}
\usepackage{amstext}
\usepackage{graphicx}

\begin{document}

\title{Light Transmission Through Metallic-Mean Quasiperiodic Stacks with Oblique Incidence}

\author{\normalsize Stefanie Thiem$^{\rm a,b}$\thanks{Corresponding author. Email: stefanie.thiem@physik.tu-chemnitz.de}$\;$, Michael Schreiber$^{\rm a}$, and Uwe Grimm$^{\rm c}$\\
\small$^{\rm a}${\em{Institut f\"ur Physik, Technische Universit\"at Chemnitz, D-09107 Chemnitz, Germany}}\\
\small$^{\rm b}${\em{Department of Physics, University of California, Berkeley, CA 94720, USA}}\\
\small$^{\rm c}${\em{Department of Mathematics and Statistics, The Open University, Milton Keynes, MK7 6AA, UK}}}

\date{}
\maketitle

\begin{abstract}\vspace{-1cm}
\small The propagation of s- and p-polarized light through quasiperiodic multilayers, consisting of layers with different refractive indices, is studied by the transfer matrix method. In particular, we focus on the transmission coefficient of the systems in dependency on the incidence angle and on the ratio of the refractive indices.
We obtain additional bands with almost complete transmission in the quasiperiodic systems at frequencies in the range of the photonic band gap of a system with a periodic alignment of the two materials for both types of light polarization. With increasing incidence angle these bands bend towards higher frequencies, where the curvature of the transmission bands in the quasiperiodic stack depends on the metallic mean of the construction rule. Additionally, in the quasiperiodic systems for p-polarized light the bands show almost complete transmission near the Brewster's angle in contrast to the results for s-polarized light.
Further, we present results for the influence of the refractive indices at the midgap frequency of the periodic stack, where the quasiperiodicity was found to be most effective.
\bigskip

\textbf{Keywords:} quasicrystals, multilayers, photonic materials, transmission coefficient
%
\end{abstract}

\section{Introduction}

The insight in the photonic properties of materials is important for the construction of customized new optical devices. In this context, quasicrystals, which are regarded to have a degree of order intermediate between crystals and disordered systems, are of special interest because they often possess unexpected physical properties due to their complex symmetries \cite{PhysRevLett.1984.Shechtman, UsefulQuasicrystals, PhysicalProperties.1999.Stadnik}. This qualifies them for the potential application in several optical devices such as single-mode light-emitting diodes, polarization switches, and  optical filters \cite{OptExp.2008.Hendrickson, PhilMag.2008.Bahabad, OptComm.1998.Garzia}. An interesting application is also the construction of microelectronic devices that are based on photons rather than on electrons, so that they potentially can be the electromagnetic analogue to semiconductors \cite{PhysWorld.2004.McGrath,JPhysD.2007.Steurer}.

The theoretical study of photonic properties of one-dimensional systems is based on the transfer matrix method and the concept of aperiodic mathematical sequences, as e.g. the Fibonacci sequence, the Thue-Morse sequence, or Cantor sequences \cite{JPhysD.2007.Steurer,PhysRevLett.1987.Kohmoto,PhysRevE.2009.Esaki,JPhys.2007.Yin}. Such one-dimensional systems can be relatively easily produced in reality and a comparison of the theoretical and the experimental results shows a good agreement \cite{PhysRevLett.1994.Gellermann,JPhys.2009.Nava}.

While in literature most results are only given for zero incidence angle, s-polarized light and fixed refractive indices, we compare the transmission of s- and p-polarized light through different multilayers in dependency on the incidence angle of the light and the refractive indices.

\begin{figure}
 \centering
 \includegraphics[width=10cm]{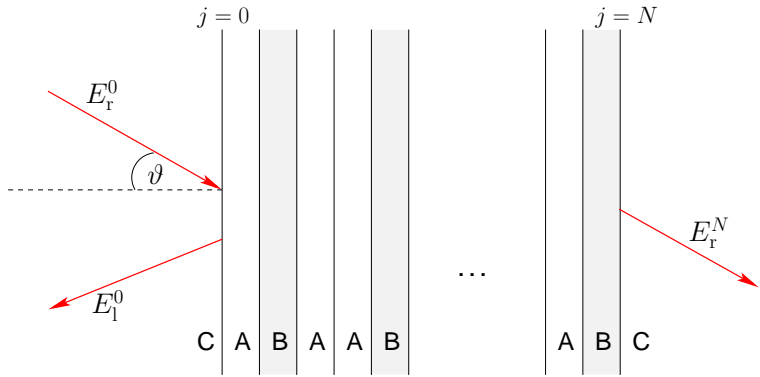}
 \caption{Transmission of light through a quasiperiodic stack consisting of the materials $A$ and $B$ with different refractive indices. The light wave $E_\mathrm{r}^0$ is incident on the surface under an angle $\vartheta$. Parts of the light wave are transmitted through the stack ($E_\mathrm{r}^N$) and other parts are reflected ($E_\mathrm{l}^0$).}
 \label{fig:geometry}
\end{figure}

In particular, the configuration of the layers of the so-called metallic-mean quasicrystals can be constructed by the inflation rule
 \begin{equation}
  \label{equ:infaltion_rule}
  \mathcal{P}_a =
  \begin{cases}
    B \rightarrow A \\
    A \rightarrow A^a B
  \end{cases}\hspace{1cm},
 \end{equation}
iterated $m$ times starting from the symbol $B$. We refer to the resulting sequence after $m$ iterations as the $m$th-order approximant $\mathcal{C}_m^a$ with the length $f_m$ given by the recursive equation $f_m = f_{m-2} + a f_{m-1}$ with the initial values $f_0 = f_1 = 1$. Depending on the parameter $a$ the inflation rule generates different metallic means, i.e., the lengths of two successive sequences satisfy the relation
 \begin{equation}
  \label{equ:tau}
  \lim_{m \rightarrow \infty} \frac{f_m}{f_{m-1}} = \lim_{m \rightarrow \infty} \frac{f_m^A}{f_m^B} = \tau(a) \;,
 \end{equation}
where $\tau(a)$ is an irrational number with the continued-fraction representation $\tau(a) = [\bar{a}] = [a,a,a,...]$. Also, the number $f_m^A$ of layers $A$ and number $f_m^B$ of layers $B$ in an approximant $\mathcal{C}_m^a$ are related by this metallic mean.

For example, $a=1$ yields the well-known Fibonacci sequence related to the golden mean $\tau_\mathrm{Au} = [\bar{1}] = (1+\sqrt{5})/2$,
the case $a=2$ corresponds to the octonacci sequence with the silver mean $\tau_\mathrm{Ag} = [\bar{2}] = 1+\sqrt{2}$, and for $a=3$ one obtains the bronze mean $\tau_\mathrm{Bz} = [\bar{3}] = (3+\sqrt{13})/2$. In general the relation $\tau(a) = (a+\sqrt{a^2+4})/2$ holds \cite{NonLinAnal.1999.Spinadel, JPhys.2007.Yin}.

The outline of this paper is as follows: In Sec.~\ref{sec:transfer_matrix} we give an introduction to the transfer matrix method used for the calculations and in Sec.~\ref{sec:results} we present our results and discuss them. This is followed by a brief conclusion in Sec.~\ref{sec:conclusion}.

\section{Transfer Matrix Method}\label{sec:transfer_matrix}

\begin{figure}[h!]
 \centering
 \includegraphics[width=5cm]{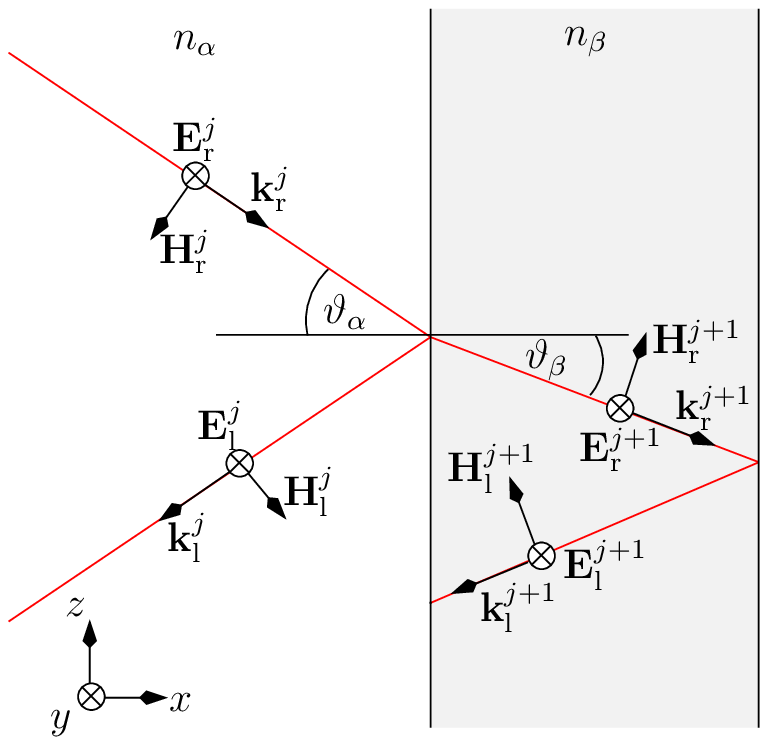}\hspace{2cm}
 \includegraphics[width=5cm]{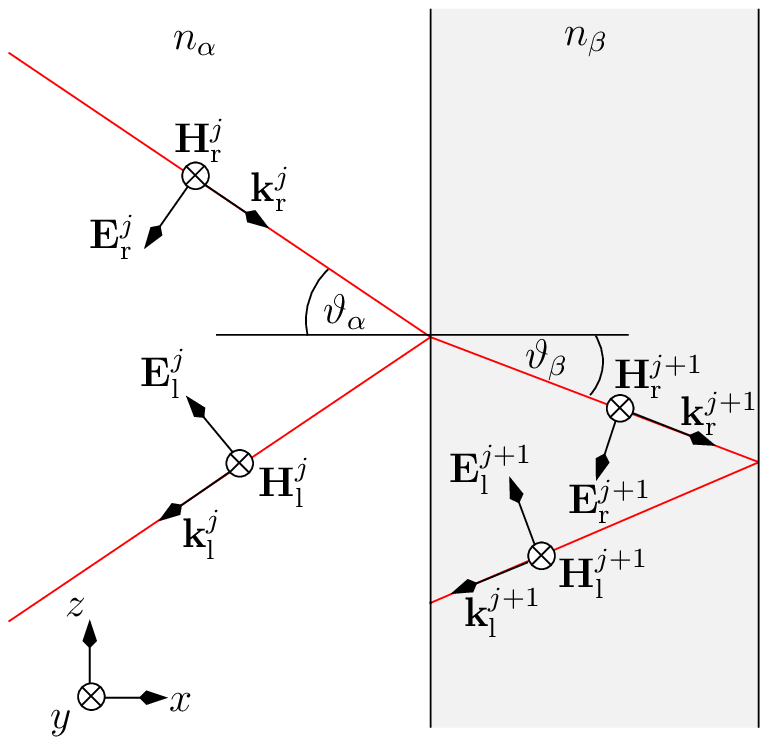}
 \caption{Electric and magnetic field vectors $\mathbf{E}$ and $\mathbf{H}$ at the interface between layer $j$ and $j+1$ with different refractive indices $n_\alpha$ and $n_\beta$ for s-polarized light (left) and p-polarized light (right).}
 \label{fig:geometry2}
\end{figure}

The propagation of light through a layered system as shown in Fig.~\ref{fig:geometry} is commonly investigated by transfer matrix methods \cite{PhysRevLett.1987.Kohmoto,PhysRevE.2009.Esaki,JPhys.1998.Vasconcelos,JPhys.2006.deMedeiros}. Regarding the geometry we compare the propagation of linearly polarized light with the electric field perpendicular to the plane of the light path (s-polarization) and with the electric field vector lying in the plane of the light path (p-polarization) (cp.~Fig.~\ref{fig:geometry2}) \cite{Book.Hecht}.

Within one layer the electric field $\mathbf{E} = E \mathbf{\hat{e}}_y$ can be described as a superposition of a right and a left traveling plane wave
 $E = E_\mathrm{r}^j e^{ik_jx - i\omega t} + E_\mathrm{l}^j e^{-ik_jx - i\omega t},$
where $k_j = n_j k$ denotes the wave number of the light in the $j$th layer with refractive index $n_j$, $k$ the wave number in the vacuum, and $\omega$ the frequency of the light.

The propagation of the waves through the stack is given on the one hand by their refraction at the interfaces between the layers and on the other hand by their propagation through the layers. Addressing the first point, the boundary conditions at the interfaces between the layers require the tangential component of the electric field $\mathbf{E}$ and the magnetic field $\mathbf{H} = \tfrac{n}{c} \mathbf{\hat{k}} \times \mathbf{E}$ to be continuous across the boundaries \cite{OpticalWaves.1998.Yeh}. Applying the $(E_+, E_-)^\mathrm{T}$notation used by Kohmoto et al. \cite{PhysRevLett.1987.Kohmoto,PhysRevE.2009.Esaki} with
 $E_+ = E_\mathrm{r} + E_\mathrm{l} \text{  and  } E_- = -i\left(E_\mathrm{r} - E_\mathrm{l}\right),$
the interface matrices $T_{\alpha\beta} =  T_{\beta\alpha}^{-1}$ for s- and p-polarized light, respectively,
\begin{align}
 T_{\alpha\beta}^\mathrm{s} &=
 \begin{pmatrix}
  1 & 0 \\
  0 & \tfrac{n_{\beta} \cos{\vartheta_{\beta}}}{n_{\alpha} \cos{\vartheta_{\alpha}}}
 \end{pmatrix}&
 T_{\alpha\beta}^\mathrm{p} &=
 \begin{pmatrix}
  1 & 0 \\
  0 & \tfrac{n_{\beta} \cos{\vartheta_{\alpha}}}{n_{\alpha} \cos{\vartheta_{\beta}}}
 \end{pmatrix}
\end{align}
describe the transfer of the light wave from medium $\alpha$ to medium $\beta$.
Thereby, the incidence angle $\vartheta_{\alpha}$, the emergence angle $\vartheta_{\beta}$, and the corresponding refractive indices $n_\alpha$ and $n_\beta$ are related by Snell's law according to
$ n_{\alpha} \sin{\vartheta_{\alpha}} = n_{\beta} \sin{\vartheta_{\beta}} $. Further, total reflection occurs at incidence angles $\vartheta > \vartheta_{\mathrm{total}} = \arcsin{(n_B/n_A)}$.

On the other hand, the propagation of the light waves within the layers results in a phase difference, which is comprised in the propagation matrix
\begin{equation}
 T_{\gamma} =
 \begin{pmatrix}
  \cos{\varphi_{\gamma}}& \sin{\varphi_{\gamma}} \\
  -\sin{\varphi_{\gamma}} & \cos{\varphi_{\gamma}}
 \end{pmatrix}\;.
\end{equation}
The phase difference for the wave length $\lambda = 2\pi / k$ of the light and a layer thickness $d_\gamma$ amounts to $ \varphi_{\gamma} = 2 k d_{\gamma} n_{\gamma}\cos{\vartheta_{\gamma}}$ \cite{Book.Hecht}.
For the choice of the wave length of the incident light $\lambda_0 = 2\pi c /\omega_0 $, we use the commonly applied quarter wave length condition $n_A d_A = n_B d_B = \lambda_0/4$, which leads for an incidence angle $\vartheta = 0$ to an identical optical wave path of the light in the two materials $A$ and $B$ of the stack \cite{JPhys.2006.deMedeiros,PhysRevLett.1994.Gellermann,JPhys.2009.Nava}. Hence, we obtain
\begin{equation}
 \label{equ:phase2}
 \varphi_{\gamma} = \frac{\lambda_0}{\lambda} \pi \cos{\vartheta_{\gamma}} = \frac{\omega}{\omega_0}\pi \cos{\vartheta_{\gamma}} \;.
\end{equation}

The transfer matrix of the overall system relates the incident light $E_\mathrm{r}^{0}$, the reflected light $E_\mathrm{l}^{0}$, and the transmitted light $E_\mathrm{r}^{N}$ (cp.~Fig.~\ref{fig:geometry}) by
\begin{equation}
 \label{equ:transfer_equation}
 \begin{pmatrix}
  E_+ \\ E_-
 \end{pmatrix}_0 = M_m^{\mathcal{C}^a}
 \begin{pmatrix}
  E_+ \\ E_-
 \end{pmatrix}_N
\end{equation}
and is given as a combination of the different interface and propagation matrices according to a certain quasiperiodic sequence.
In general, the recursive equation
\begin{equation}
 M_m^{\mathcal{C}^a} = \{ M_{m-1}^{\mathcal{C}^a} \}^{a} M_{m-2}^{\mathcal{C}^a}
\end{equation}
is applicable for $m \ge 2$ with $M_0^{\mathcal{C}^a} = T_{AB} T_B T_{BA} $ and $M_1^{\mathcal{C}^a} = T_A $. For refractive indices of the surrounding medium $n_C \ne n_A$ the corresponding transfer matrices $T_{CA}$ and $T_{AC}$ have to be added at the edges.

In particular, we are interested in the calculation of the transmission coefficient $T$ (also known as transmittance) of the light through the stack defined as
\begin{equation}
 T = \frac{|E_\mathrm{r}^N|^2 }{ |E_\mathrm{r}^0|^2} = \frac{4 (\det{M})^2}{|M|^2 + 2 \det{M}} \;,
\end{equation}
where $|M|^2$ corresponds to the sum of the squares of all four matrix elements of $M$. Further, for the complete stack it can be shown that $\det{M} = 1$.

\section{Results}\label{sec:results}

In this section we comprise the results for several metallic-mean quasicrystals. Thereby, we focus on the change of the transmission coefficient $T$ in dependence on the inflation rule $\mathcal{P}$, the light polarization, the ratio of the refractive indices $u = n_A/n_B$, and the incidence angle $\vartheta$. Here we only show results for one system size $f(\mathcal{C}^a)$. In general, increasing the number of layers results in a reduction of the width of the transmission bands and the occurrence of new transmission bands in a selfsimilar manner \cite{PhysRevLett.1987.Kohmoto,EPJB.2010.Thiem}. In the limit of infinite systems the transmission bands form a Cantor set with Lebesgue measure 0, and it is assumed that either only complete transmission or complete reflectance occurs \cite{PhysRevLett.1987.Kohmoto,PhysRevE.2009.Esaki}. However, for physical applications usually systems with relatively small numbers of layers are used \cite{OptComm.1998.Garzia, OptExp.2008.Hendrickson}.

\begin{figure*}
 \centering
 \includegraphics[height=5.3cm]{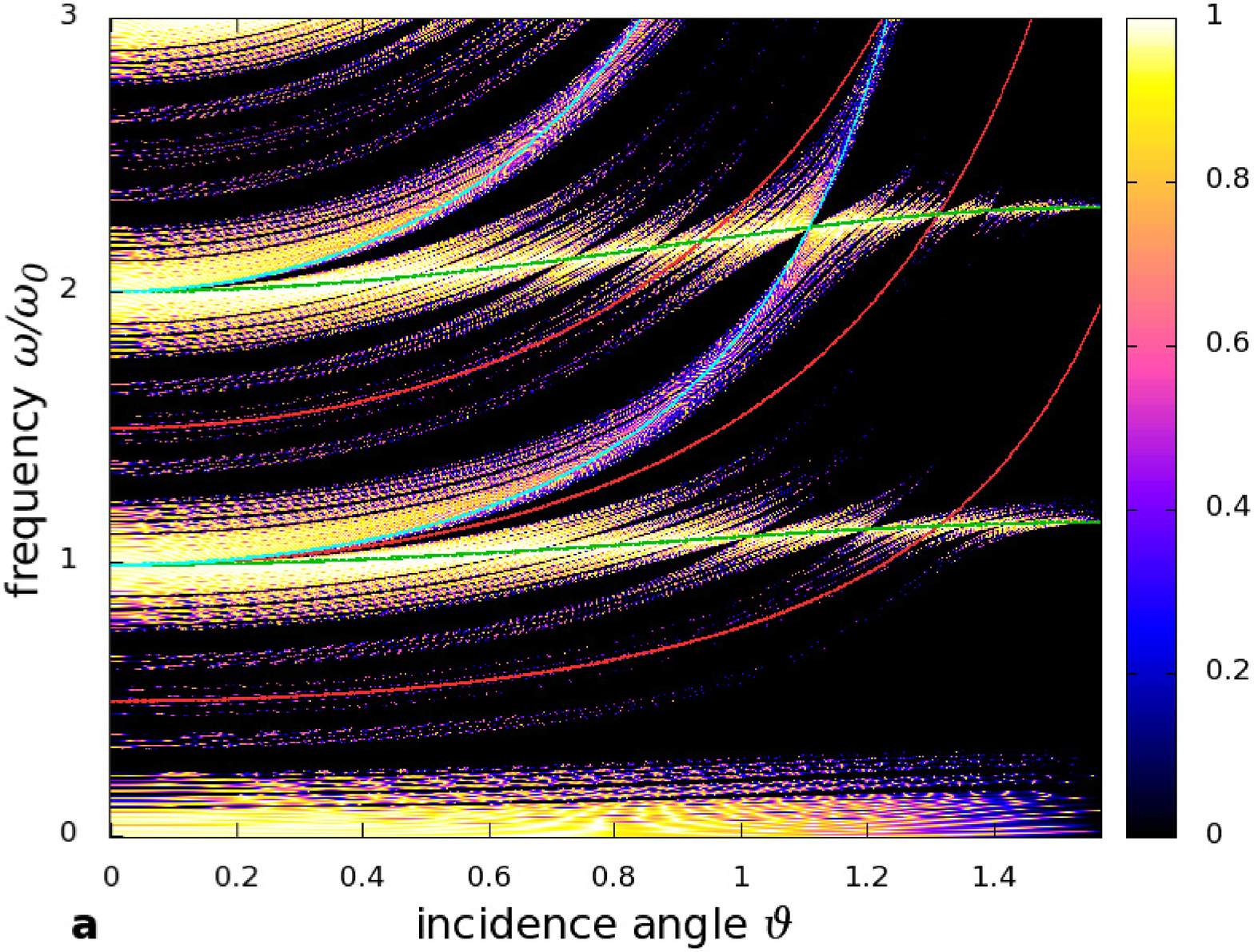}\hspace{0.5cm}
 \includegraphics[height=5.3cm]{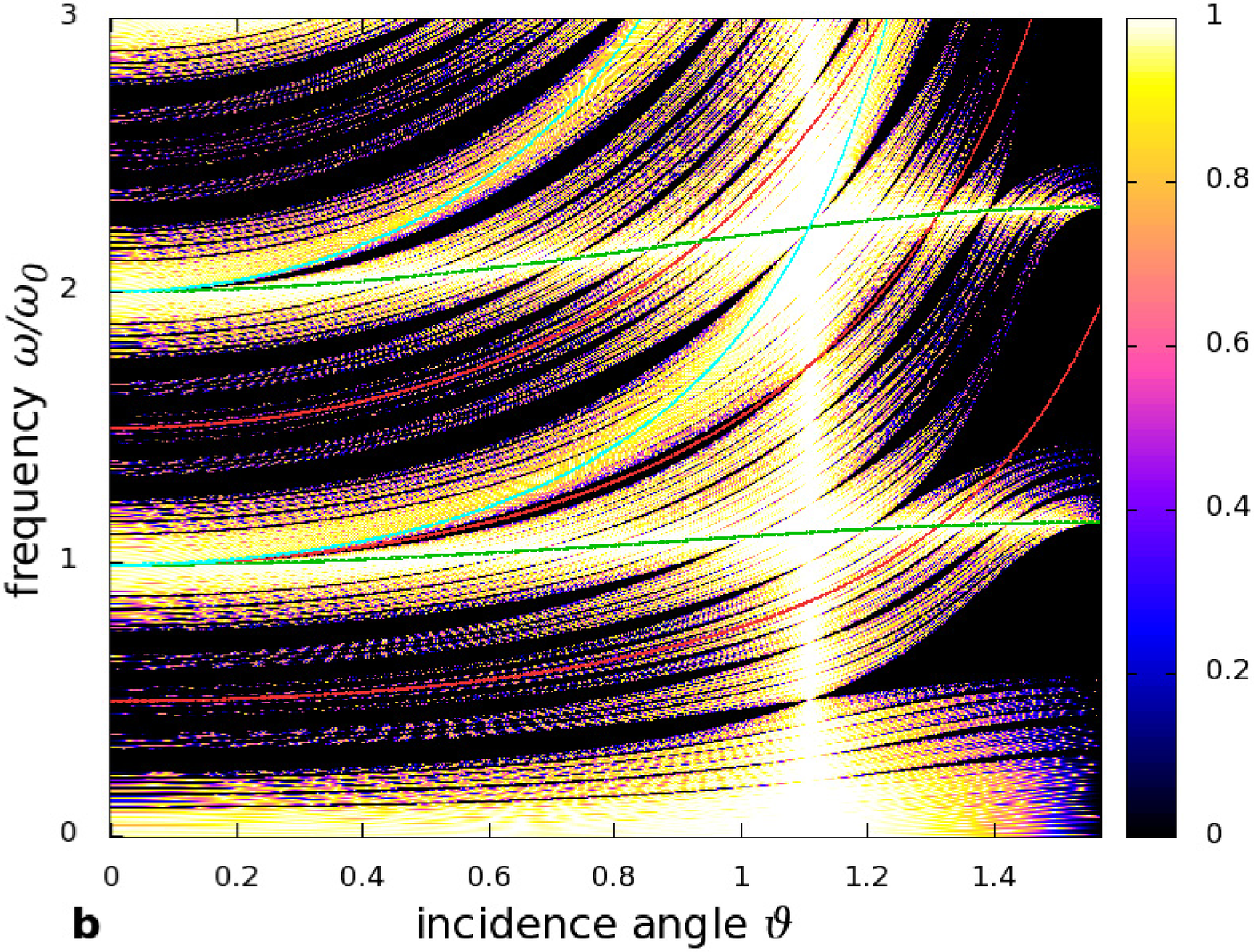}
 \caption{Transmittance $T(\vartheta, \omega)$ of light through a stack of the materials A and B, which are arranged according to the octonacci sequence $\mathcal{C}_{8}^{\mathrm{Ag}}$ with $f_8 = 577$ layers. Results are shown in dependence on the reduced frequency $\omega / \omega_0$ versus the incidence angle $\vartheta$ for $n_A = n_C$, $u = 0.5$ and (a) s-polarized and (b) p-polarized light. The blue and red lines indicate the expected bending of the transmission bands according to Eqs. \eqref{equ:bendingT2} and \eqref{equ:bendingT4}. The green lines show the regions of constructive interference according to Eq. \eqref{equ:interference}. }
 \label{fig:T_const_u}
\end{figure*}

In Fig.~\ref{fig:T_const_u} the results for the transmission coefficients $T(\vartheta, \omega)$ are shown in dependence on the reduced frequency $\omega / \omega_0$ and the incidence angle $\vartheta$ for s-polarized and p-polarized light. For $\vartheta = 0$ the transmission spectrum is periodic with respect to the frequency. However, for all angles $\vartheta > 0$ the photonic transmission bands bend towards higher frequencies for increasing incidence angles and this effect becomes stronger for higher frequencies. The main difference of the two light polarizations occurs in the region around the Brewster's angle $\vartheta_\textrm{Br} = \arctan{u^{-1}}$, which corresponds to an incidence that produces an angle of $90^\circ$ between the reflected and refracted ray. For p-polarized light we obtain in this region much higher transmission coefficients and the different transmission bands emerge more clearly even for large incidence angle, where we hardly see any transmission in the s-polarized case. Of course, for incidence angles $\vartheta = \vartheta_\textrm{Br} \simeq 1.107$ one always obtains a transmission coefficient $T = 1$ for $u = 0.5$ and p-polarized light.

\begin{figure*}
 \centering
 \includegraphics[height=5.3cm]{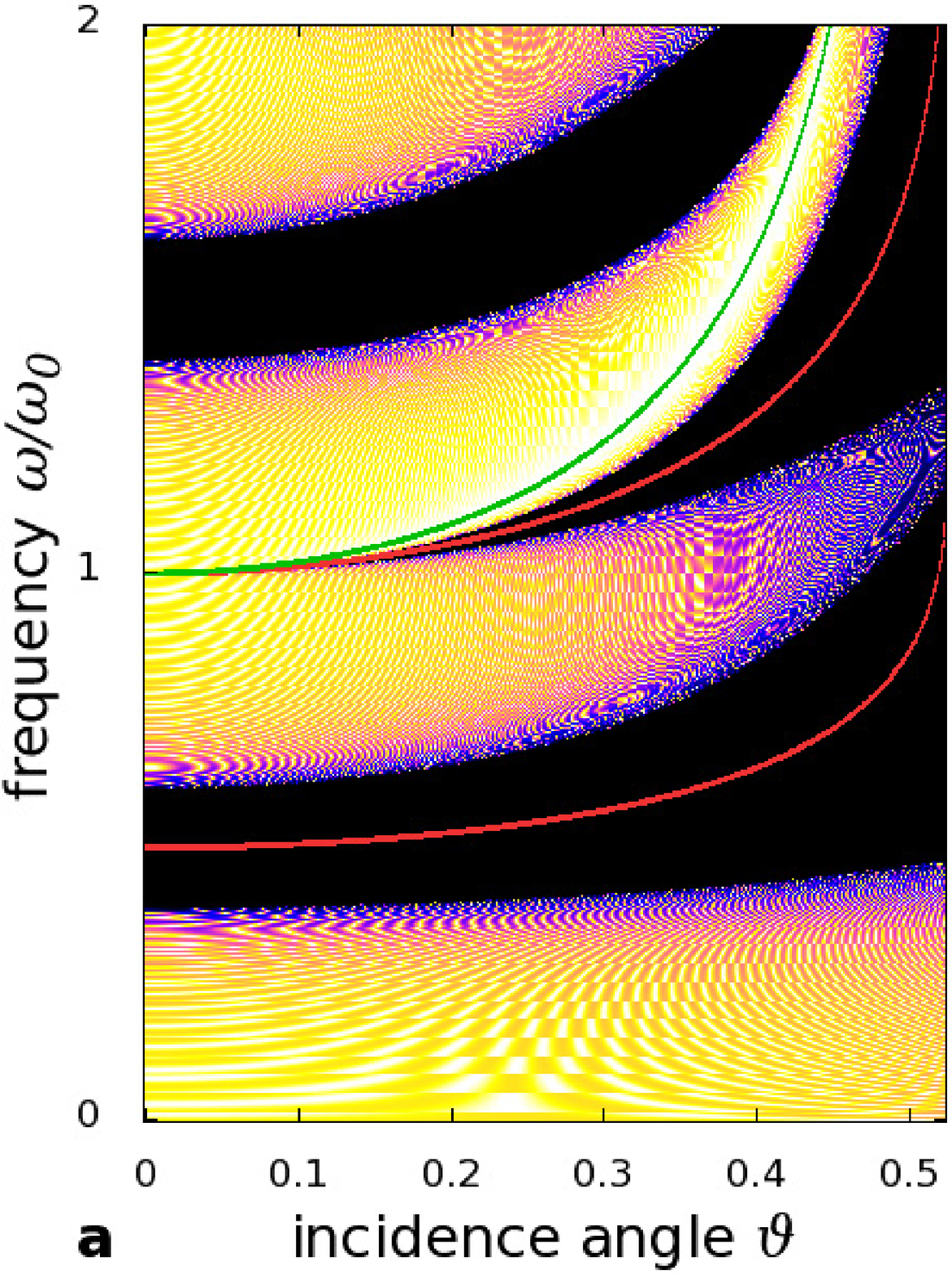}
 \includegraphics[height=5.3cm]{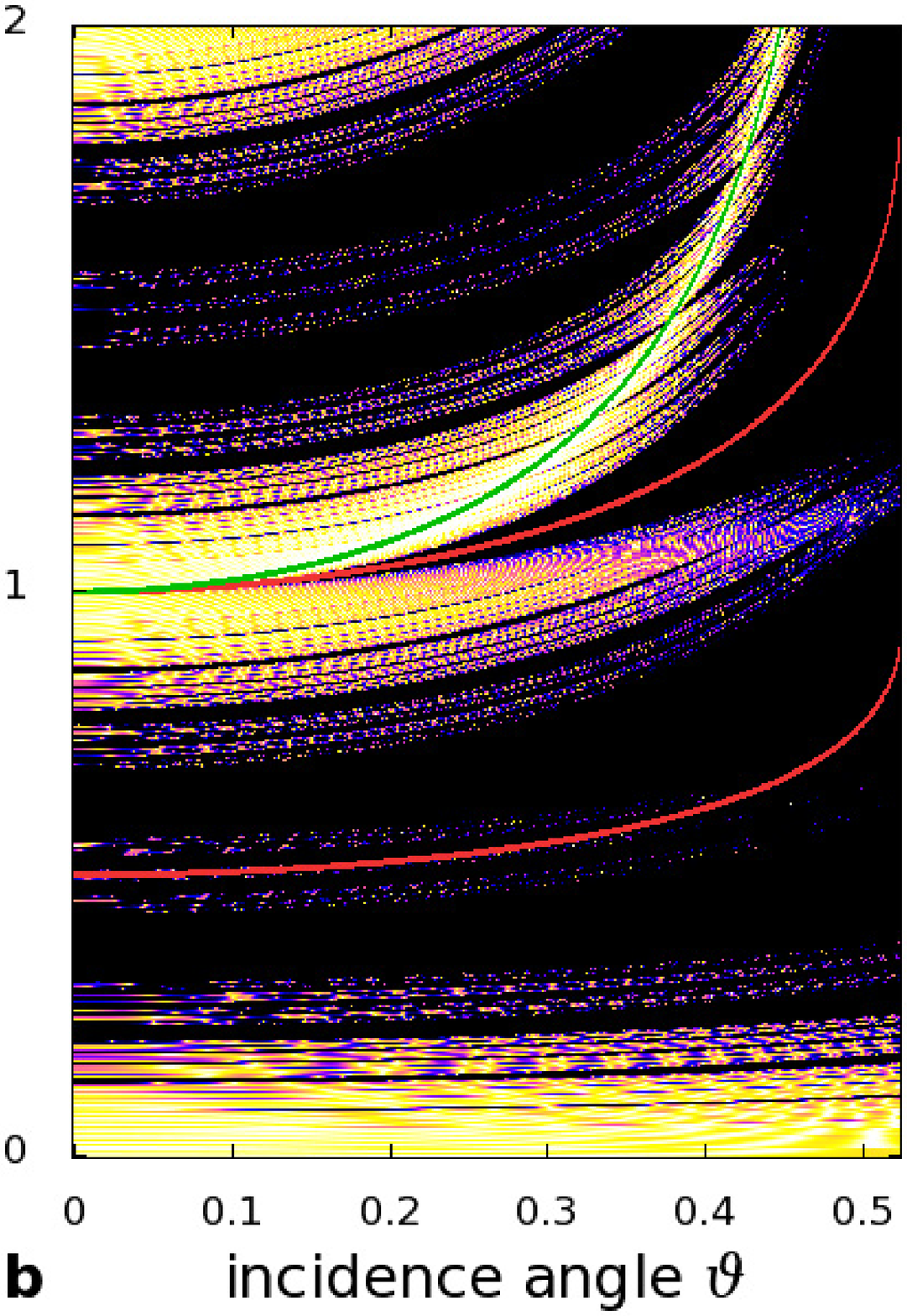}
 \includegraphics[height=5.3cm]{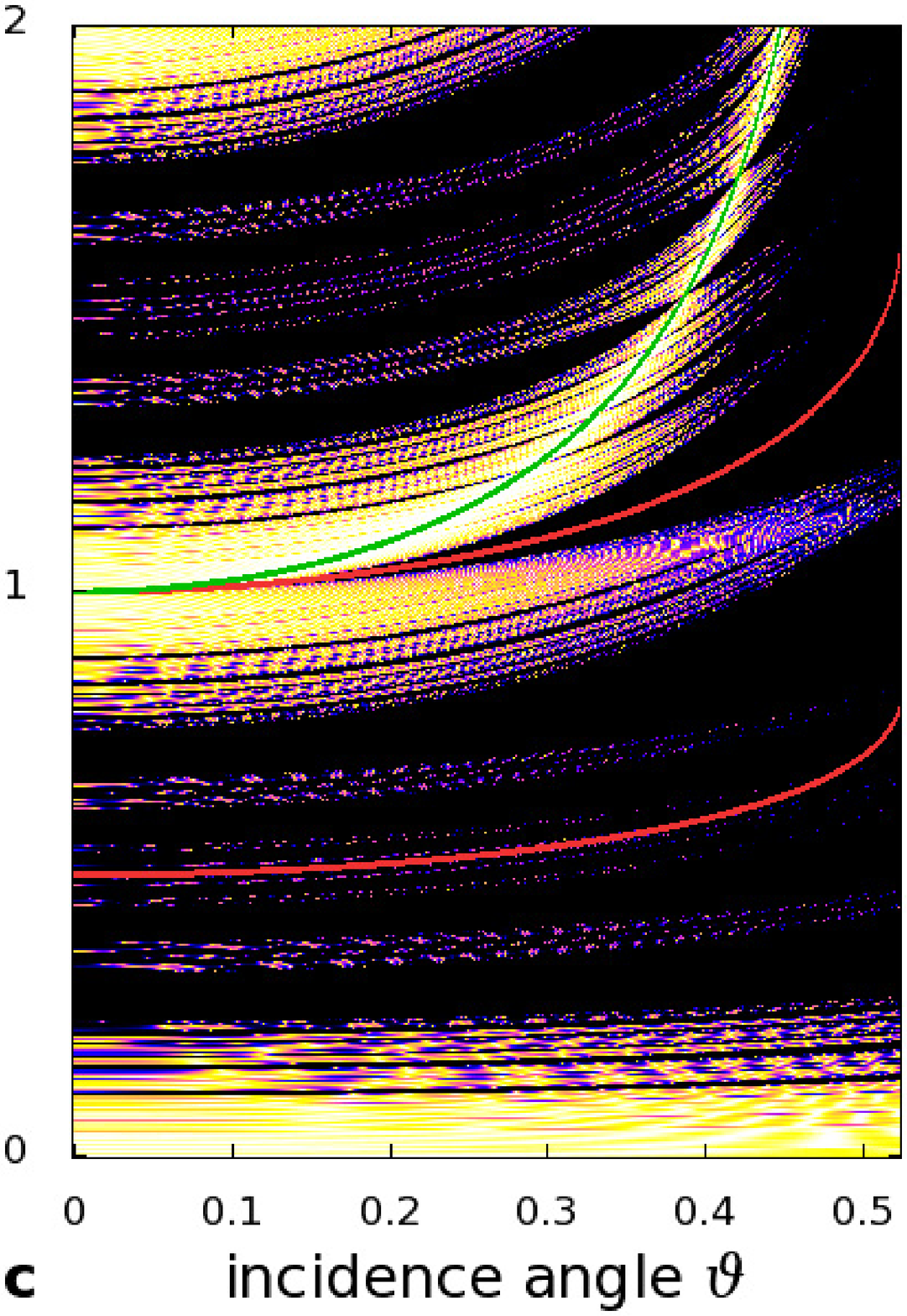}
 \includegraphics[height=5.3cm]{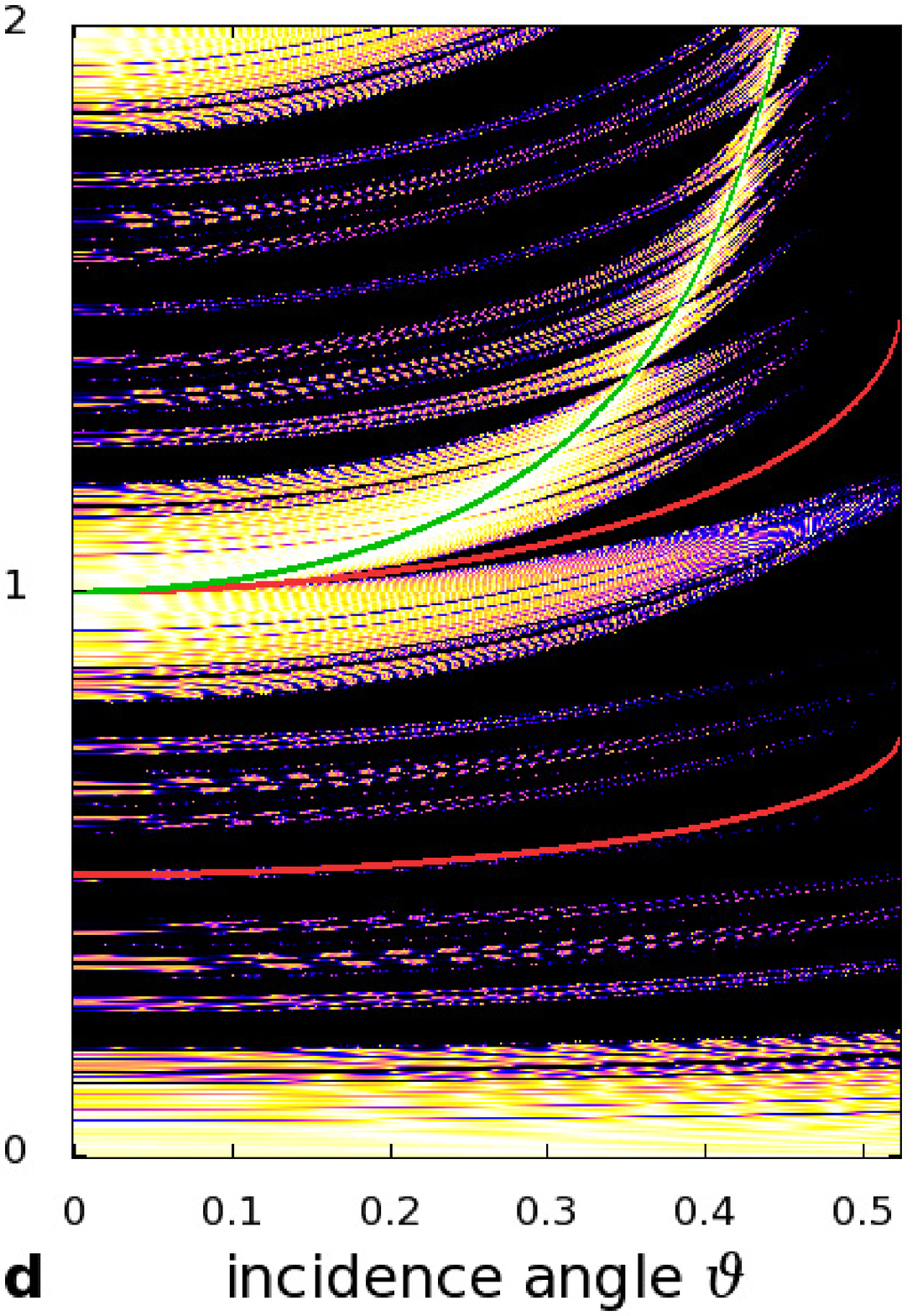}\vspace{0.2cm}
 \includegraphics[height=5.3cm]{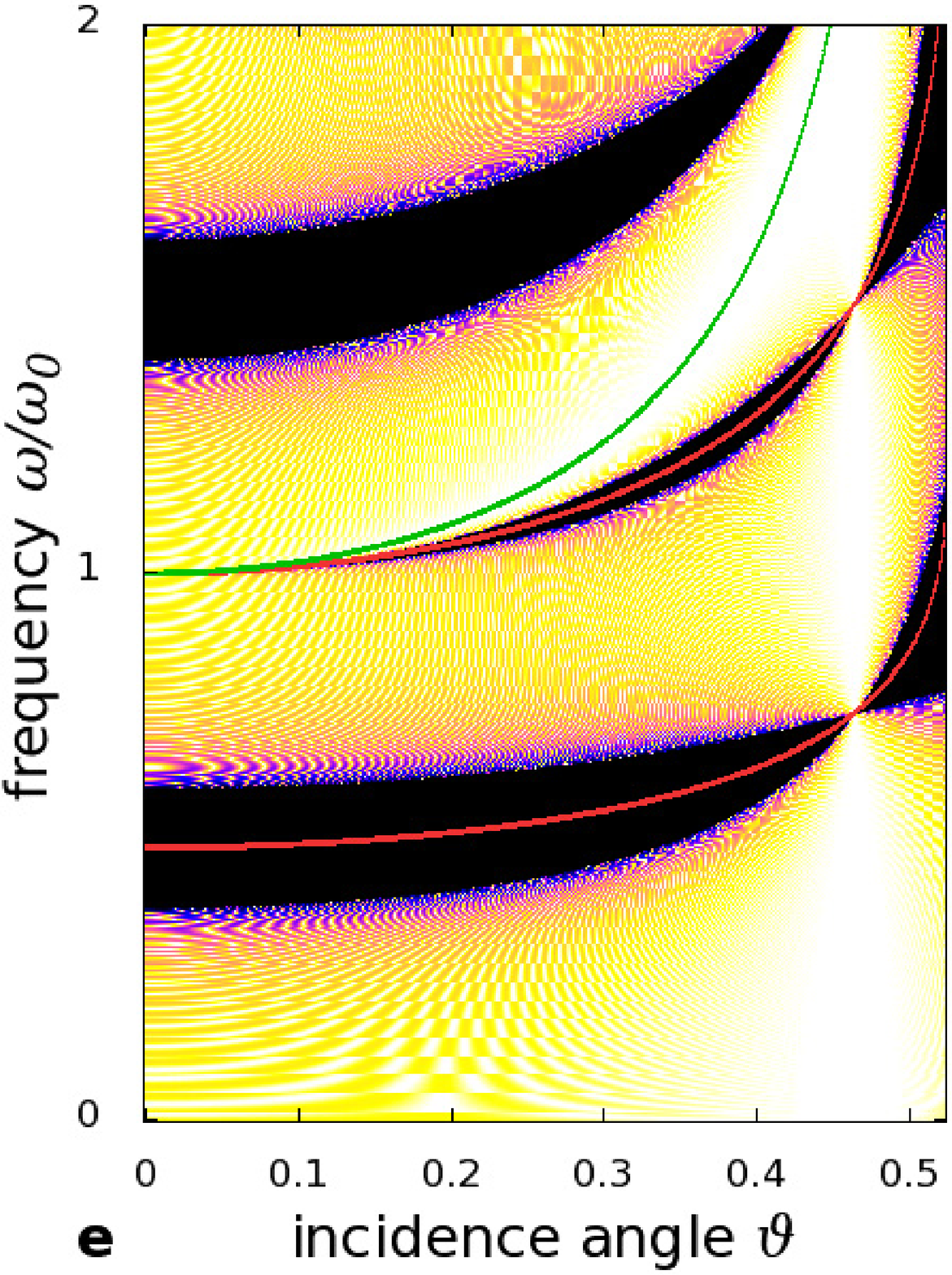}
 \includegraphics[height=5.3cm]{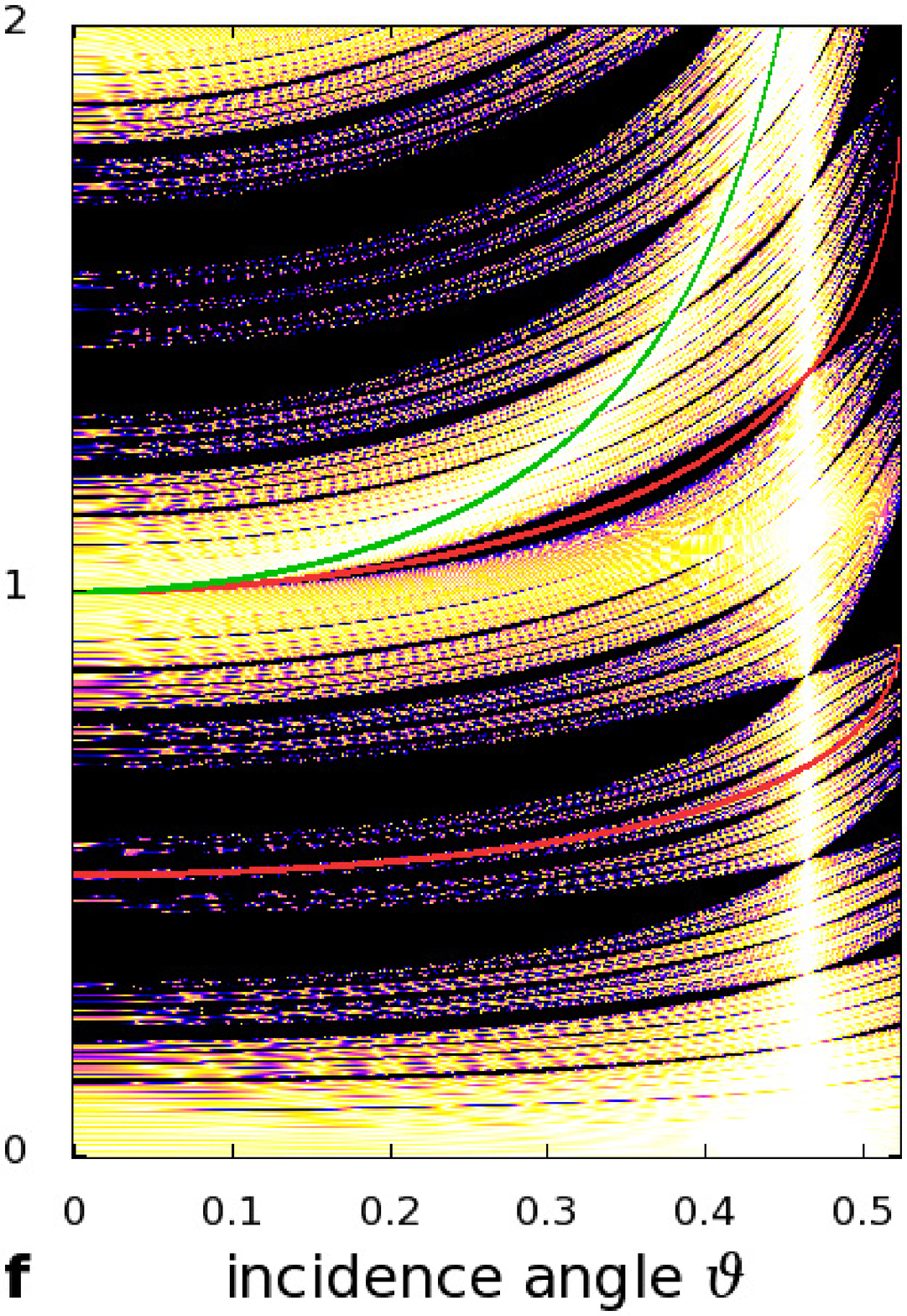}
 \includegraphics[height=5.3cm]{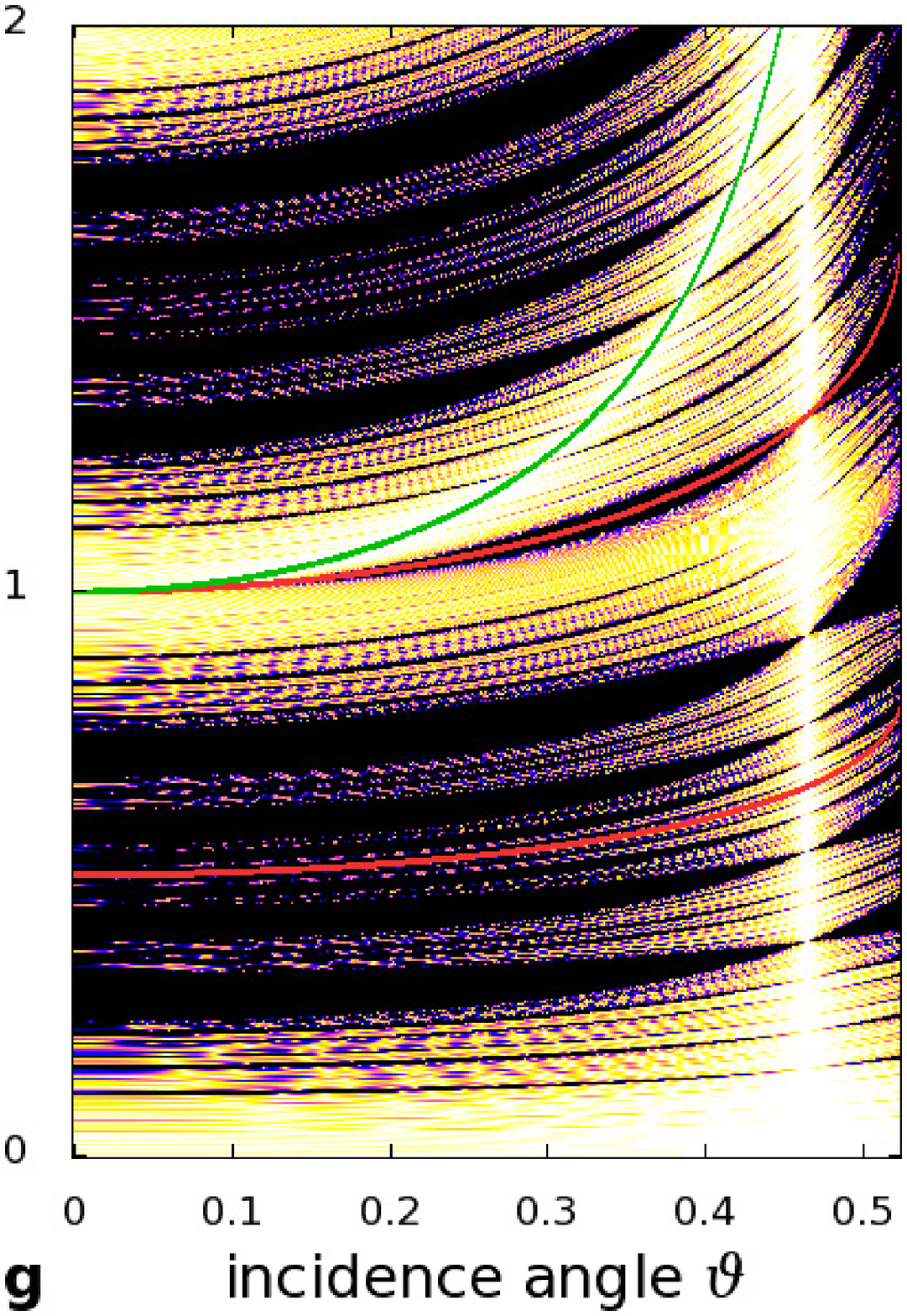}
 \includegraphics[height=5.3cm]{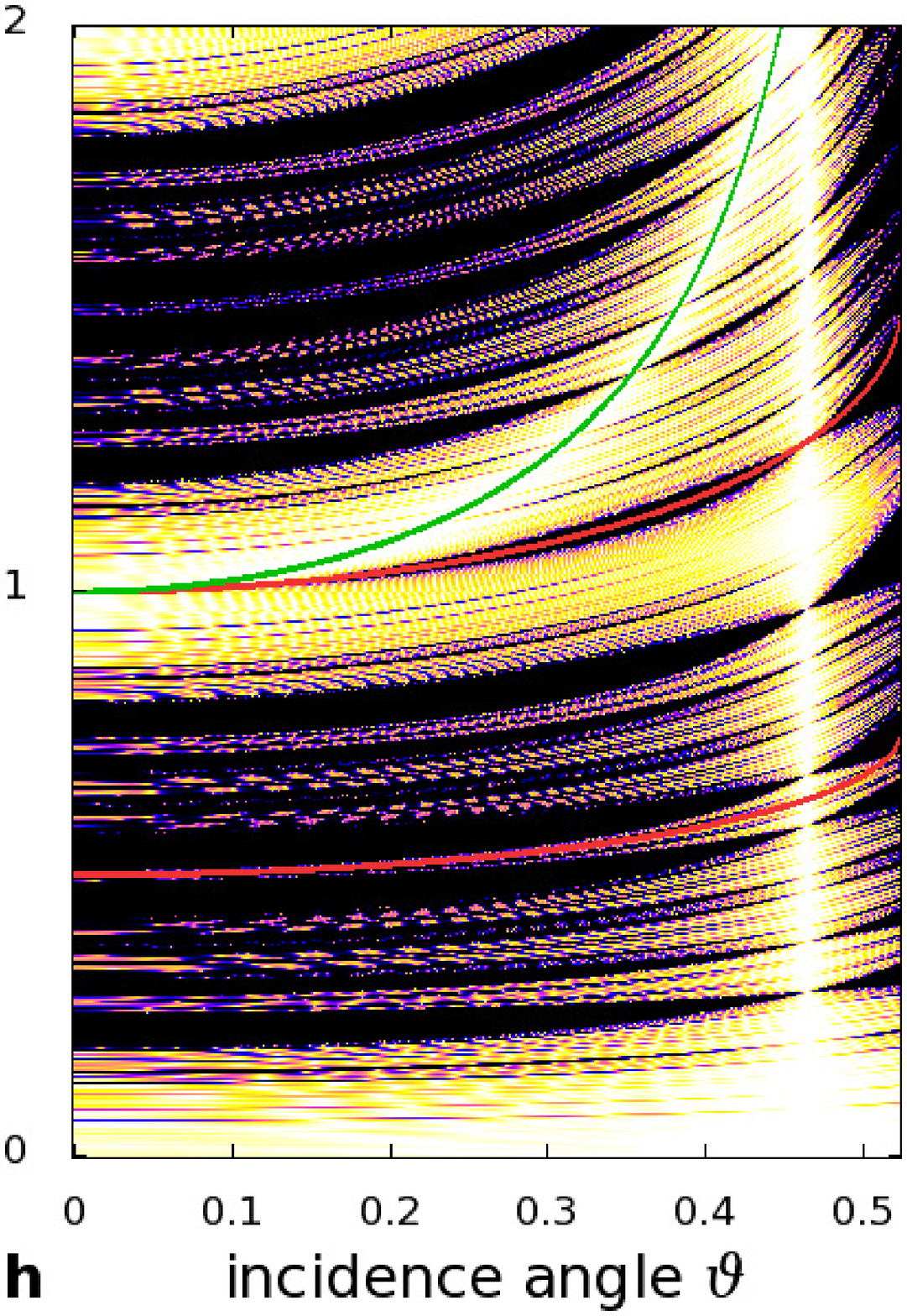}
 \caption{Transmittance $T(\vartheta, \omega)$ of different quasiperiodic stacks for s-polarized (upper row) and p-polarized light (lower row) and a ratio of the refractive indices $u=2$: (a),(e) a periodic sequence $(AB)^n$ with $f = 500$ layers, (b),(f) the Fibonacci sequence $\mathcal{C}_{14}^{\mathrm{Au}}$ with $f_ {14} = 610$ layers, (c),(g) the octonacci sequence $\mathcal{C}_8^{\mathrm{Ag}}$ with $f_8 = 577$ layers and (d),(h) the bronze-mean quasiperiodic sequence $\mathcal{C}_6^{\mathrm{Br}}$ with $f_6 = 469$ layers. The red and green lines have the same meaning as in Fig.~\ref{fig:T_const_u}.}
 \label{fig:T_all_angles}
\end{figure*}

In Fig.~\ref{fig:T_all_angles} the transmittance $T(\vartheta, \omega)$ is displayed for different metallic-mean sequences. Comparing with the results for a periodic alignment 
of the layers $A$ and $B$, one can see that there is a typical photonic band gap at $\omega / \omega_0 = b+1/2$ ($b \in \mathbb{N}_0$) for the periodic chain for $\vartheta = 0$.
For the quasiperiodic sequences this band gap appears to be much wider, but is intersected by an increasing number of new lines of moderately high transmission for increasing $a$ of the inflation rule. For $\omega / \omega_0 = b$ 
the characteristics of the transmittance $T(\vartheta, \omega)$ hardly change for the different construction rules. The behavior at the photonic band gap is in consistency with the observation that for $\omega / \omega_0 = b+1/2$ ($\vartheta = 0$) the quasiperiodicity is most effective \cite{PhysRevLett.1987.Kohmoto}. Again we observe the occurrence of almost complete transmission coefficients for p-polarized light near the Brewster's angle $\vartheta_\mathrm{Br} \simeq 0.464$.

Comparing Figs.~\ref{fig:T_const_u} and \ref{fig:T_all_angles}, i.e., the transmission for $u > 1$ and $u<1$, we observe for $u > 1$ a relatively large range of frequencies with nearly complete transmission for large incidence angles and for $u < 1$ certain frequencies with almost complete transmission over a large range of angles. These regions correspond to constructive interference for a layer $B$ which is embedded in medium $A$, i.e., the transmission coefficient of the matrix $T_{AB}T_BT_{BA}$ yields 1 for these parameter setups. This is satisfied for $\varphi_B = \pi n$ ($n \in \mathbb{Z}$) and yields a bending according to
\begin{equation}
 \label{equ:interference}
 \frac{\omega}{\omega_0} = \frac{n}{\cos{\vartheta_B}} = \frac{n}{\cos{\arcsin(u^{-1}\sin{\vartheta})}} \;.
\end{equation}
Note that for $n_A = n_C$ the relation $\vartheta_A = \vartheta$ holds. Hence, these regions of total transmission in Figs.~\ref{fig:T_const_u} and \ref{fig:T_all_angles} bend independently of the underlying (quasiperiodic or periodic) alignment of the B layers as indicated by the green lines.

A similar interference argument can be applied to estimate the curvature of the transmission bands in the quasiperiodic stack. This behavior originates from the difference of the optical light paths in the layers $A$ and $B$ for incidence angles $\vartheta > 0$. For the derivation we assume that the transmission coefficients $T(\omega, \vartheta = 0)$ and $T(\omega^\prime, \vartheta > 0)$ yield almost the same results if the overall phase difference for the stack is equal. At first, we consider only the behavior arising from the change of the phase difference within one layer. Hence, for layer $B$ the relation
$ \varphi_B (\omega, \vartheta = 0) \stackrel{!}{=} \varphi_B (\omega^{\prime}, \vartheta_B > 0)$
has to be fulfilled, which yields a behavior according to
 \begin{equation}
  \label{equ:bendingT1}
  \omega^\prime (\vartheta) \stackrel{\eqref{equ:phase2}}{=} \frac{\omega (\vartheta = 0)}{ \cos{\vartheta_B} } = \frac{ \omega (\vartheta = 0)}{ \cos{\left(\arcsin{ \left(u \sin{\vartheta}\right)}\right)}} \;.
 \end{equation}
Likewise, the relation for layer $A$ can be derived with
 \begin{equation}
  \label{equ:bendingT2}
  \omega^\prime (\vartheta) \stackrel{\eqref{equ:phase2}}{=} \frac{\omega (\vartheta = 0)}{ \cos{\vartheta} }\;.
 \end{equation}

\begin{figure*}[t]
 \centering
 \includegraphics[height=5cm]{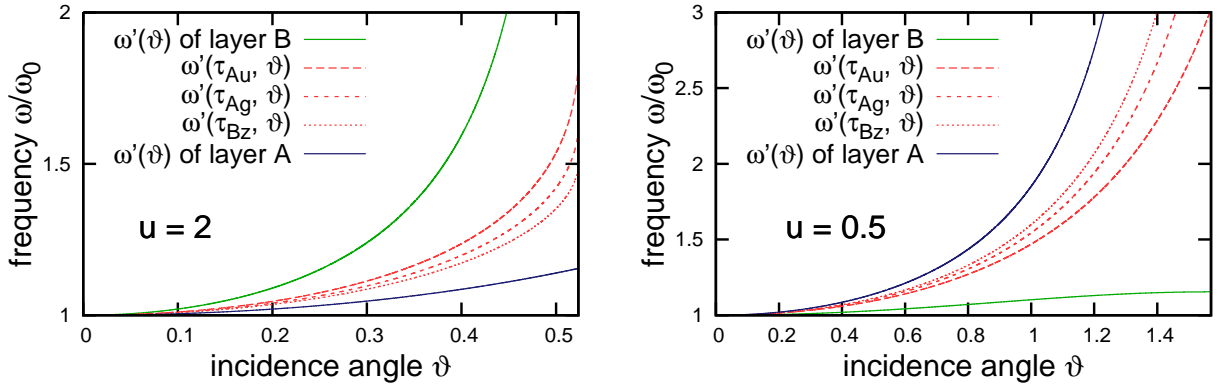}
 \caption{Limit behavior of the bending of the transmission bands with the incidence angle $\vartheta$ for quasiperiodic stacks in dependency on the metallic mean $\tau(a)$: the green, blue, and red lines correspond to Eqs. \eqref{equ:bendingT1}, \eqref{equ:bendingT2}, and \eqref{equ:bendingT4} respectively.}
 \label{fig:bendingT}
\end{figure*}

The actual bending of the transmission bands is obtained by a superposition of Eqs. \eqref{equ:bendingT1} and \eqref{equ:bendingT2}. For the quasiperiodic stacks the overall phase difference depends on the numbers of layers $A$ and $B$ in the stack according to Eq. \eqref{equ:infaltion_rule}. This results in the limit of an infinite stack in
 \begin{equation}
 \varphi = \left(\frac{\tau}{1+\tau}\varphi_A + \frac{1}{1+\tau}\varphi_B \right)f_m
 \end{equation}
and yields a bending according to
 \begin{equation}
  \label{equ:bendingT4}
  \omega^\prime (\vartheta) = \frac{(1+\tau) \omega (\vartheta = 0) }{ \tau \cos{\vartheta} + \cos{\left(\arcsin{ \left(u \sin{\vartheta}\right)}\right)} }\;.
 \end{equation}
These functions are shown in Fig.~\ref{fig:bendingT} for the cases $u=0.5$ and $u=2$. We obtain a bending between the behaviors arising from one single layer of medium $A$ or $B$. In particular, the bending increases with the parameter $\tau$ (respectively $a$) in the case $u < 1$ and shows the reverse behavior for $u > 1$. Further, for the quasiperiodic systems the bending according to Eq. \eqref{equ:bendingT4} approaches for $\tau \to 1$ the bending of a periodic stack and for $\tau \gg 1$ the behavior arising from one single layer $B$.

Comparing Eq. \eqref{equ:bendingT4} with the numerical results of the transmission for the quasiperiodic stacks for different values of $\tau$ (cp. Fig.~\ref{fig:T_all_angles}), we find a very good resemblance of the bending behavior for small and intermediate angles. While this indicates that the curvature of the transmission bands is solely caused by the phase difference, which is described by the propagation matrix, the differences in the transmission intensity for s- and p-polarized light near the Brewster's angle can only be caused by the different transfer matrices.

\begin{figure*}
 \centering
 \includegraphics[height=5.3cm]{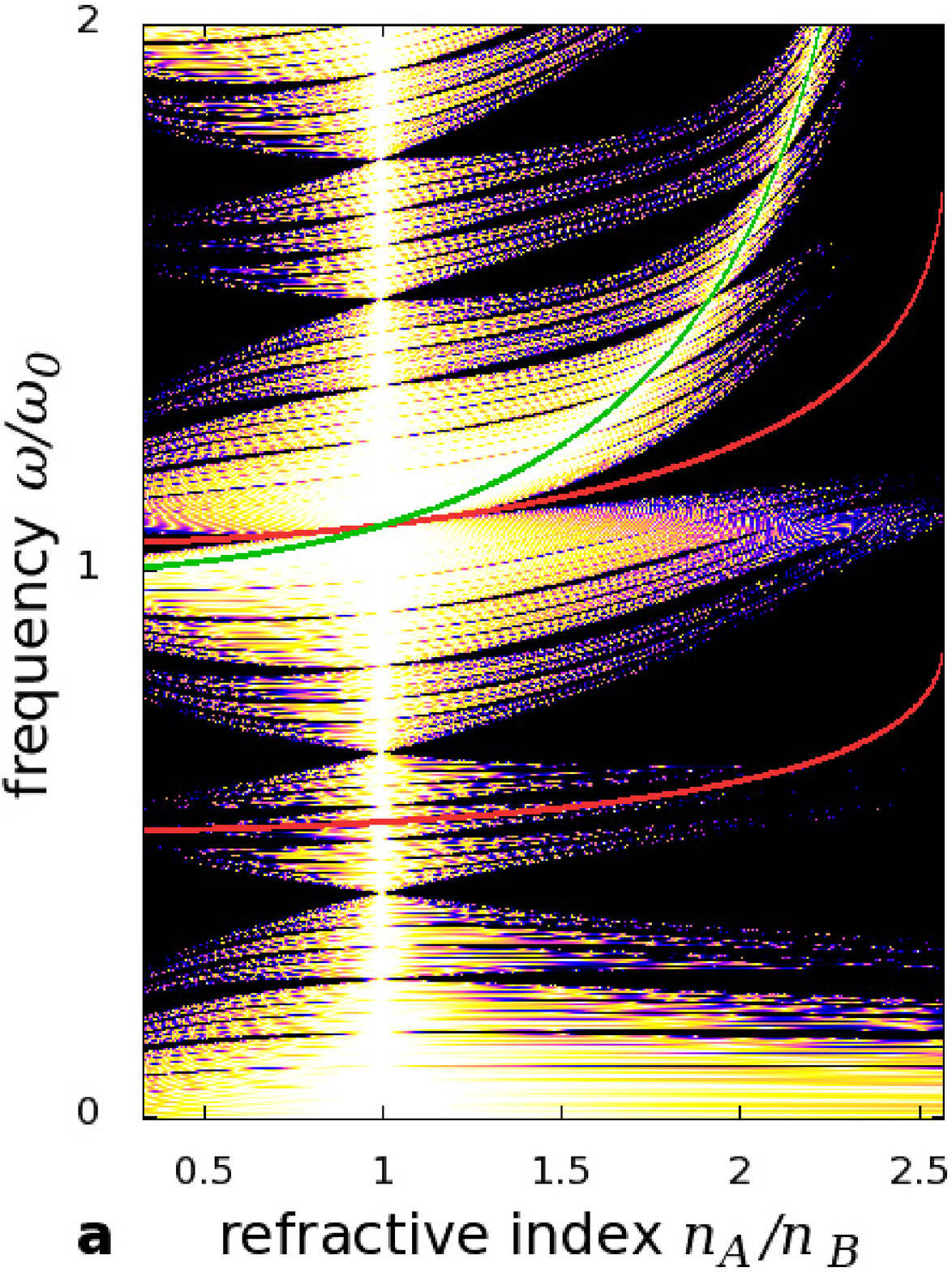}
 \includegraphics[height=5.3cm]{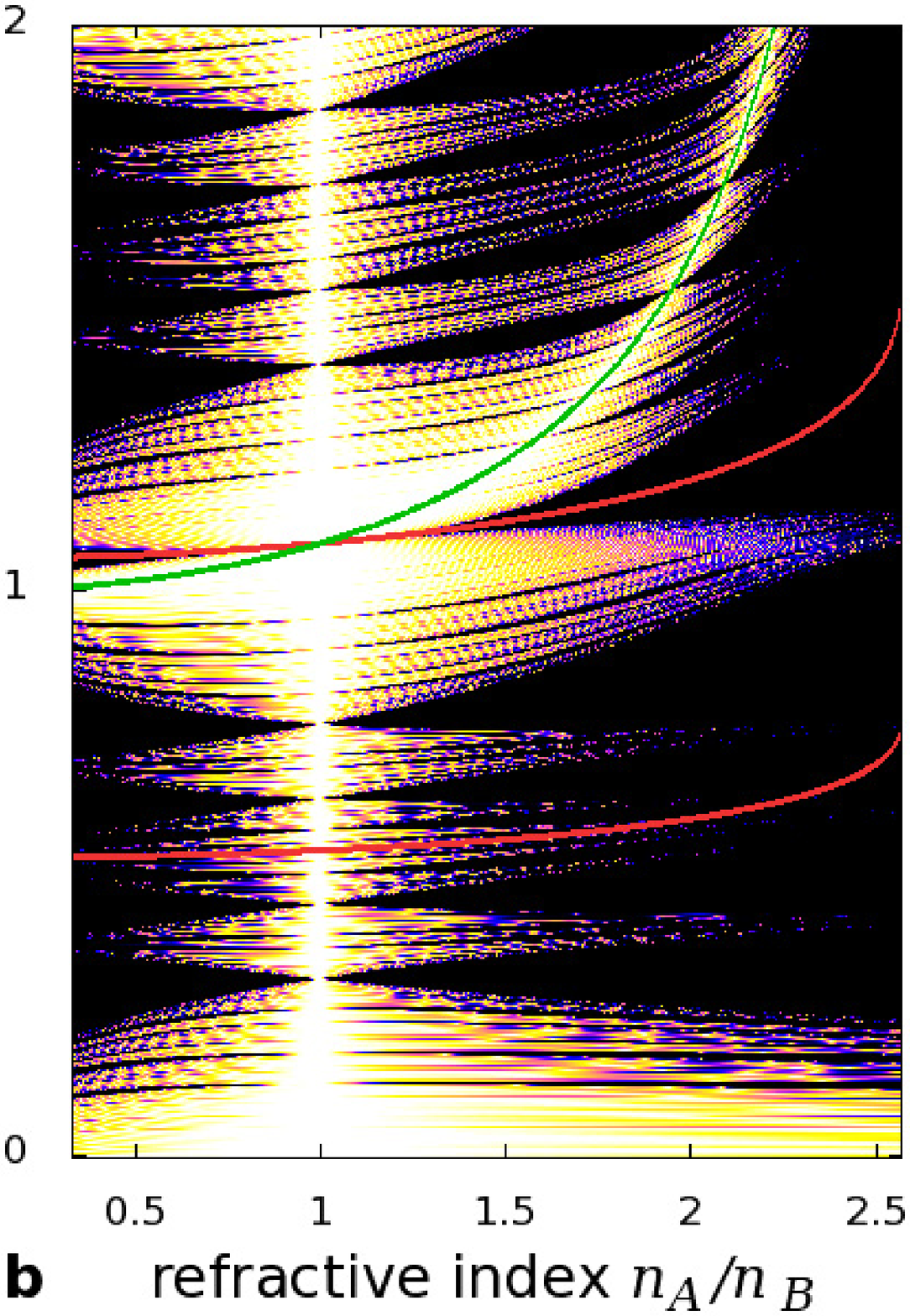}
 \includegraphics[height=5.3cm]{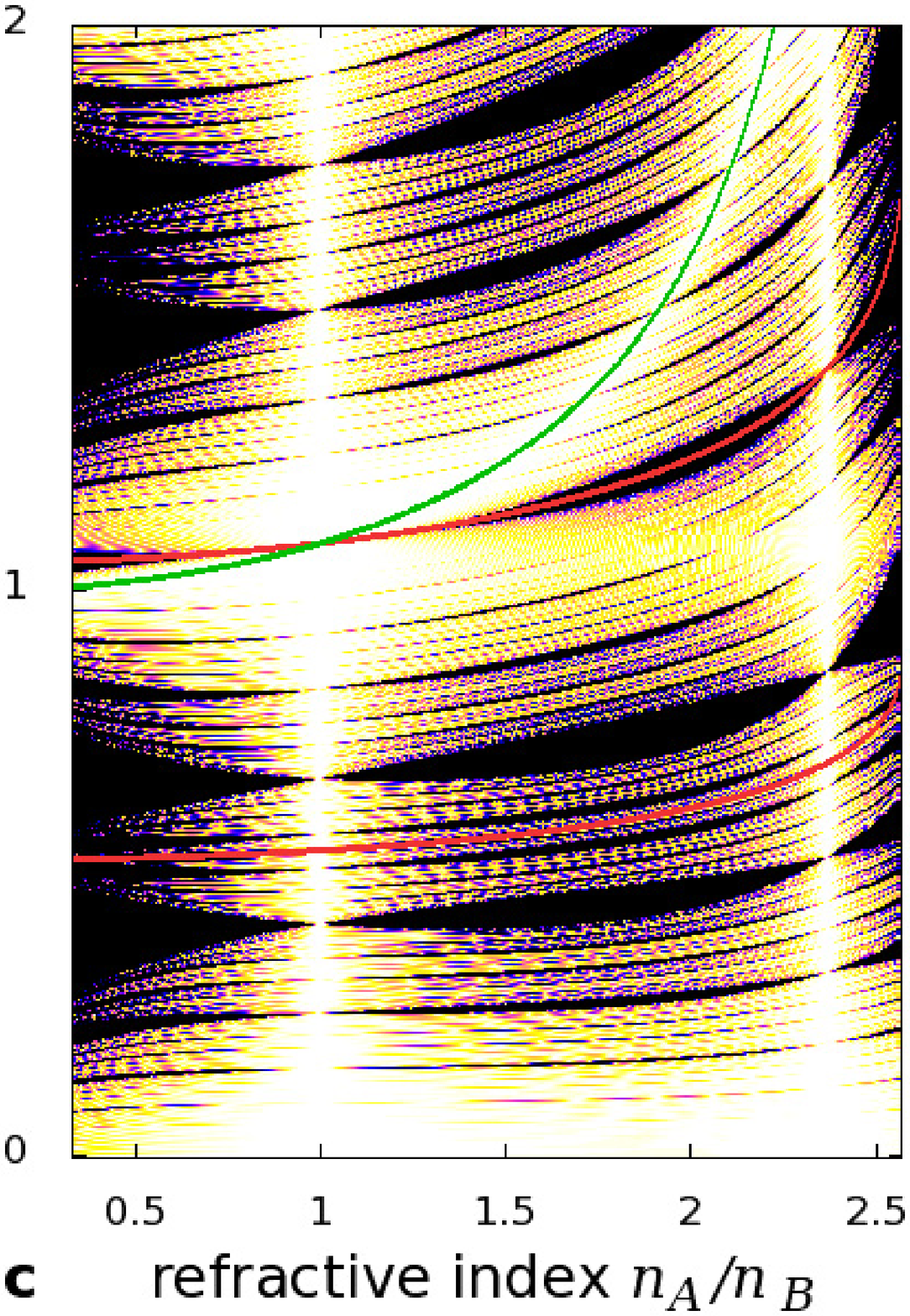}
 \includegraphics[height=5.3cm]{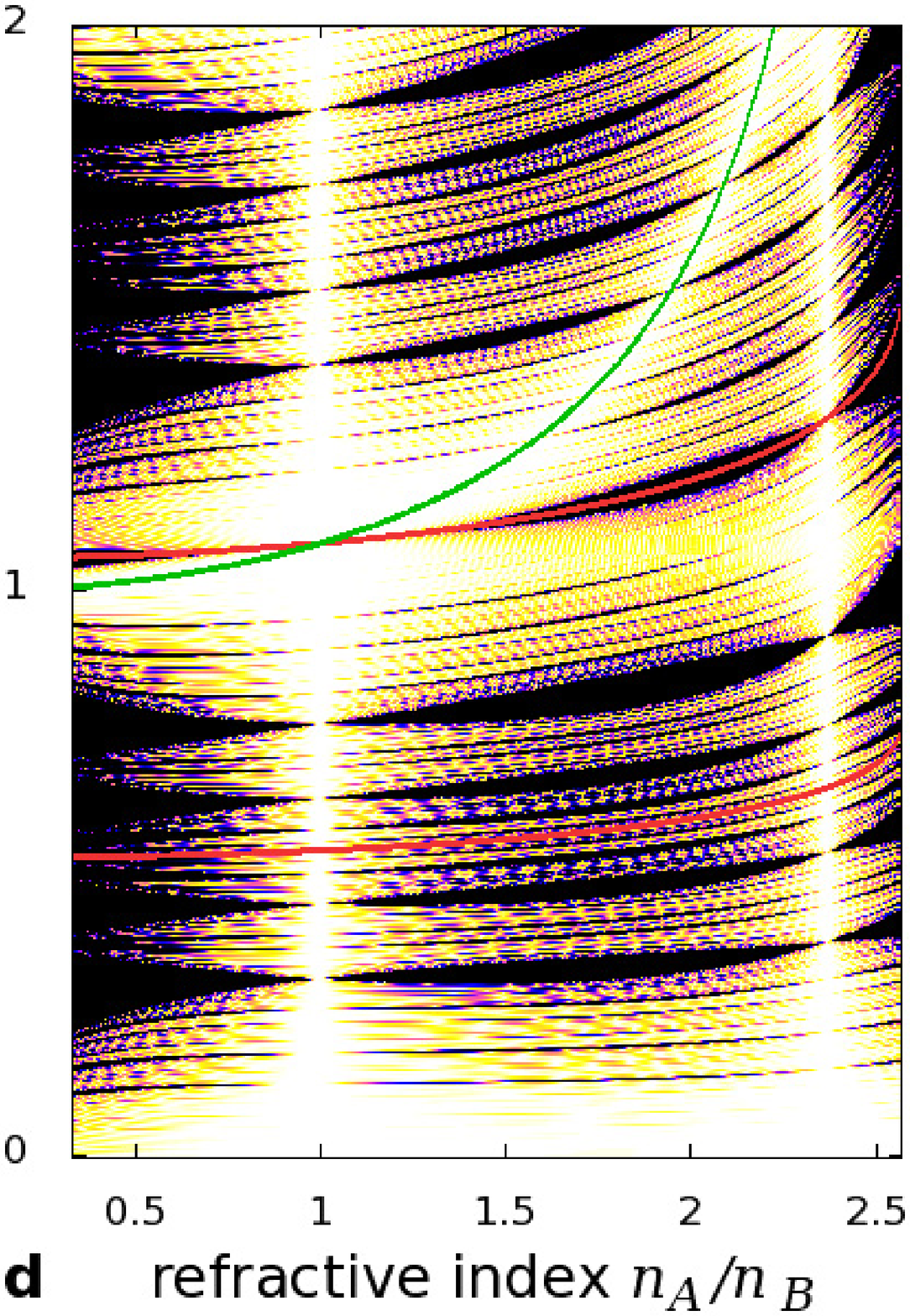}
 \caption{Transmittance $T(\omega, u)$ of quasiperiodic stacks for $\vartheta = 0.4$ and (a),(b) s-polarized and (c),(d) p-polarized light: (a),(c) the Fibonacci sequence $\mathcal{C}_{14}^{\mathrm{Au}}$ with $f_ {14} = 610$ layers and (b),(d) the octonacci sequence $\mathcal{C}_8^{\mathrm{Ag}}$ with $f_8 = 577$ layers. The red and green lines have the same meaning as in Fig.~\ref{fig:T_const_u}.}
 \label{fig:T_const_frequ}
\end{figure*}

In the following, the characteristics of the transmittance $T$ are investigated with respect to the change of the ratio of the refractive indices $u$. In Fig.~\ref{fig:T_const_frequ} we study $T(\omega, u)$ in dependency on the reduced frequency $\omega / \omega_0$ and the ratio of the refractive indices for the golden and silver mean model. Again it is clearly visible that the quasiperiodicity has the largest effect for $\omega \approx \omega_0/2$. In this region the transmittance $T(\omega, u)$ strongly varies with the used construction rule. By increasing the parameter $a$, more and more bands appear and the width of the transmission bands becomes smaller. In the periodic case complete transmission occurs for this frequency only for the obvious case $u = 1$, whereas for the quasiperiodic systems there are several bands in this region with a high transmission coefficient. Again, the bending of the bands is well described by Eq. \eqref{equ:bendingT4}. Further, for p-polarized light we obtain almost complete transmission near the ratio of the refractive indices $u_\mathrm{Br} = 1/\tan{\vartheta}$ for $\vartheta=0.4$, i.e., in this case the Brewster condition is met. This leads to a wide range of values $u \in [1,  u_\mathrm{Br}]$ for which almost complete transmission can be observed.

\begin{figure*}
 \centering
 \includegraphics[height=5.3cm]{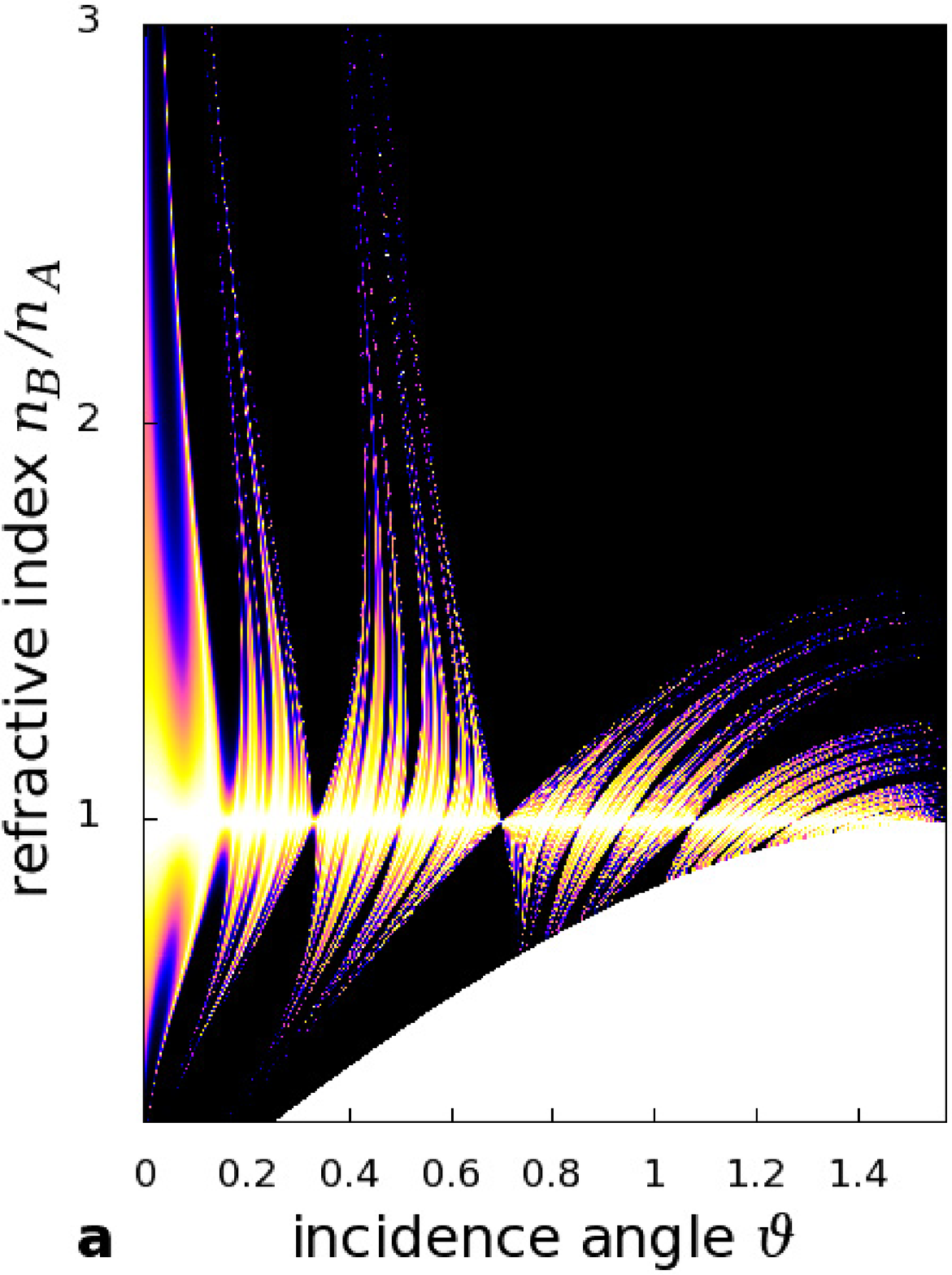}
 \includegraphics[height=5.3cm]{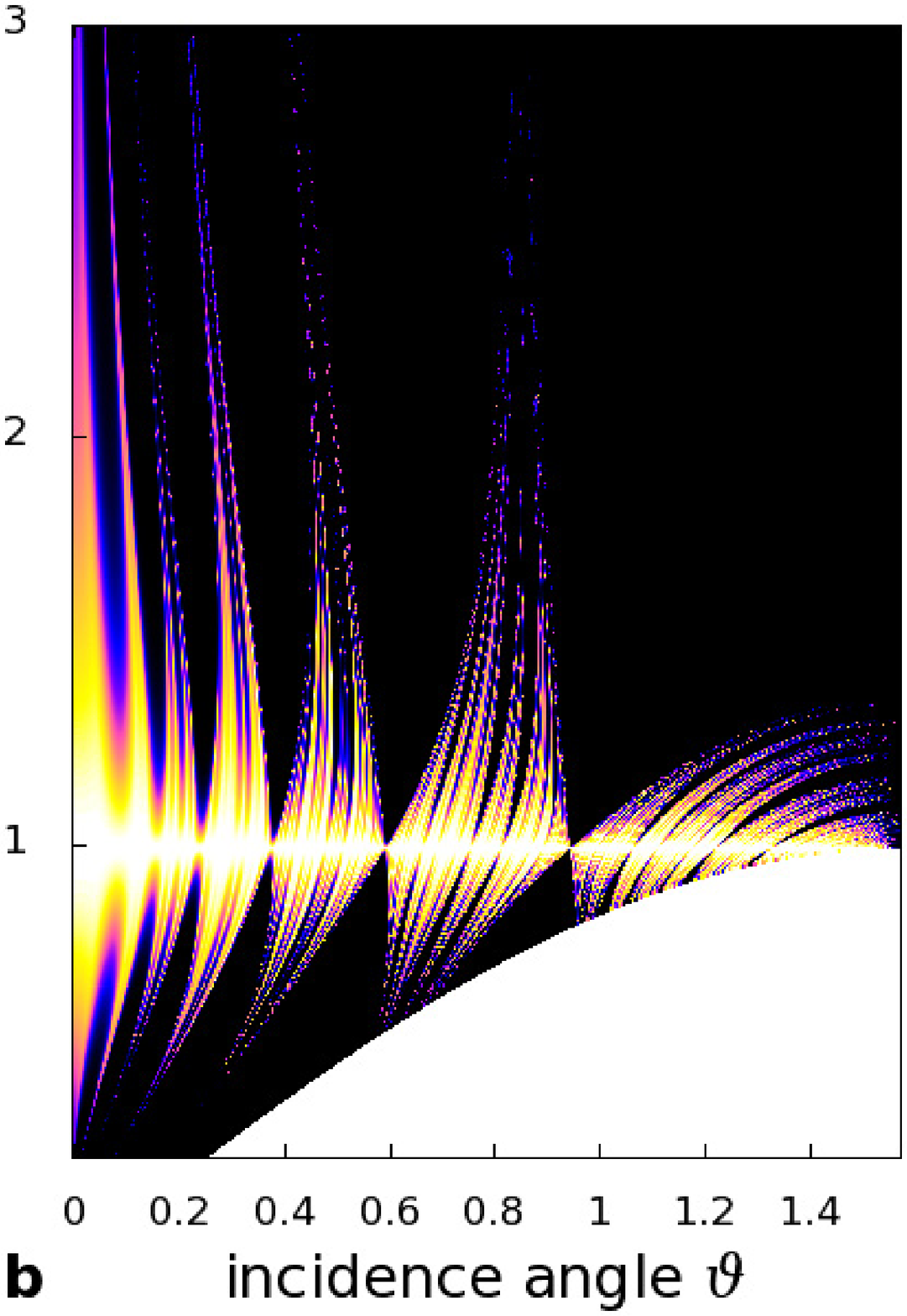}
 \includegraphics[height=5.3cm]{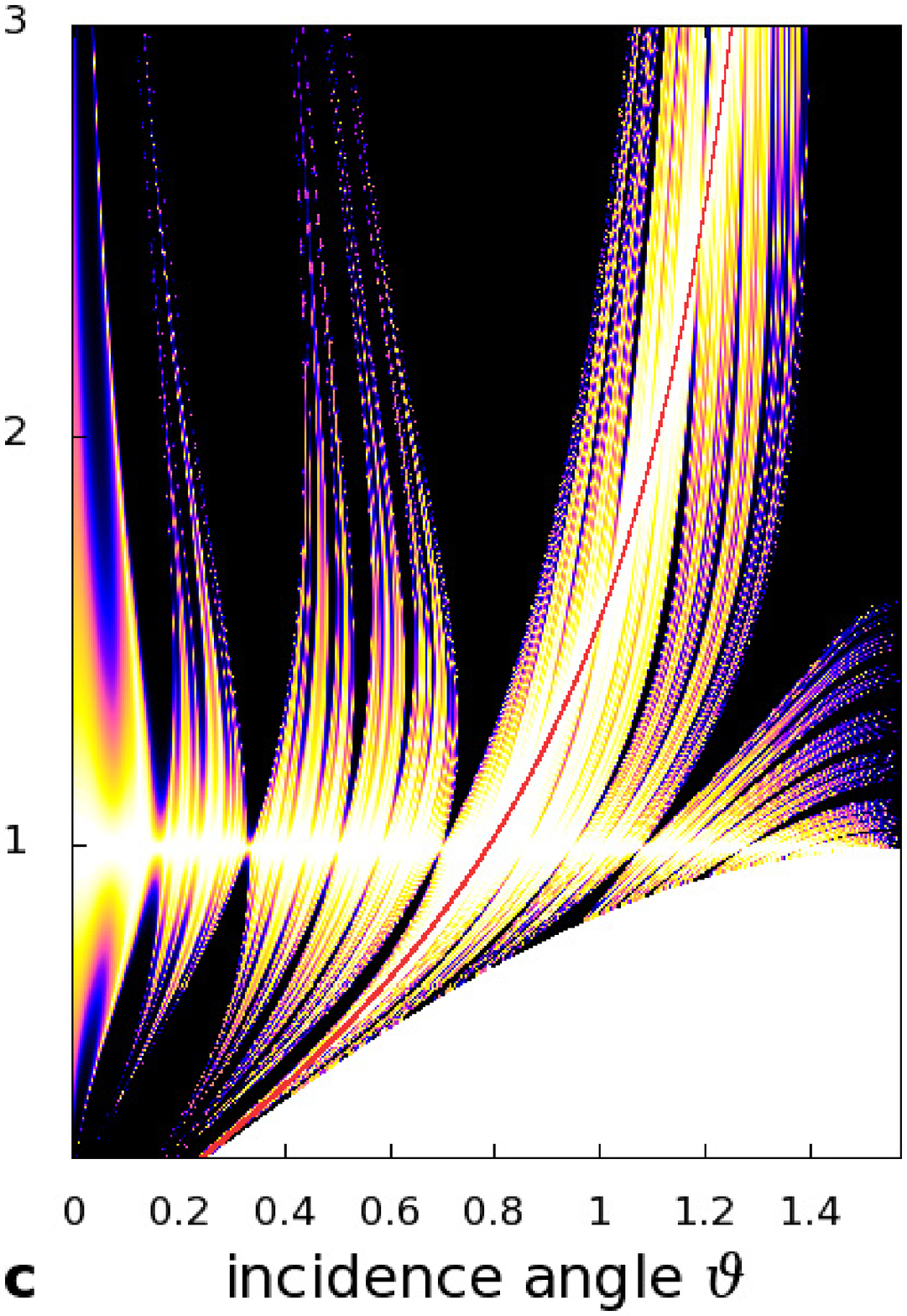}
 \includegraphics[height=5.3cm]{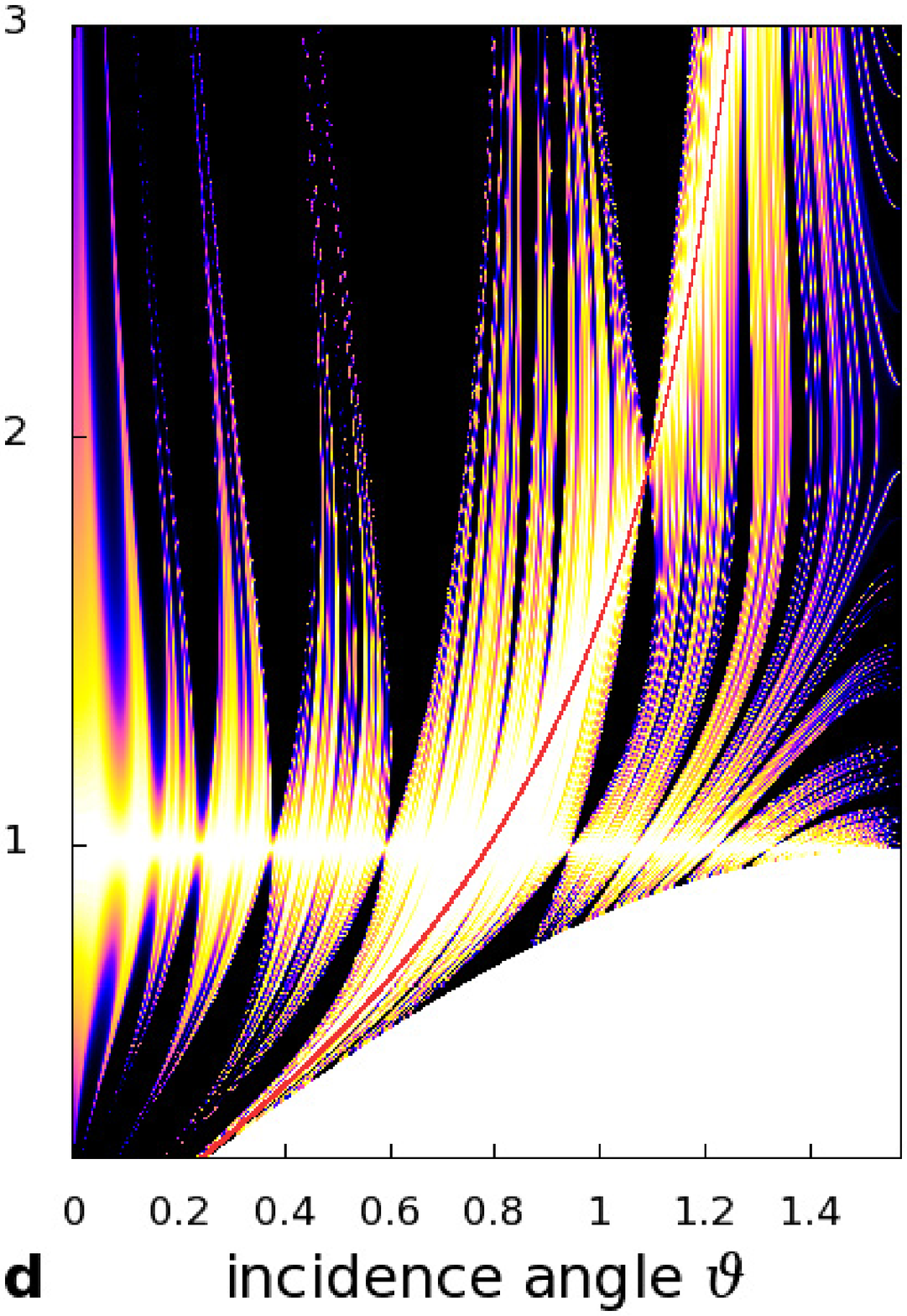}
 \caption{Transmittance $T(\vartheta, u)$ for the midgap frequency $\omega = \omega_0/2$ and the same systems as in Fig.~\ref{fig:T_const_frequ}. For parameters in the white areas in the bottom right of the plots total reflection occurs. In plots (c),(d) the Brewster condition is indicated by a red line.}
 \label{fig:T_const_angle}
\end{figure*}

The results for the influence of the refractive indices on the incidence angle $\vartheta$ are shown in Fig.~\ref{fig:T_const_angle}, where the red lines in Fig.~\ref{fig:T_const_angle}(c),(d) indicate the Brewster's angle. A comparison of the plots for s- and p-polarized light shows that the transmission is again strongly increased around the Brewster's angle. Comparing the results for the golden and silver mean model, we observe a similar structure of the transmittance $T(\vartheta,u)$ for small angles and near the regions of total reflection. While for small incidence angles transmission in a periodic system is only possible for values of $u$ close to $1$, in the considered quasiperiodic systems there is a whole range of possible ratios of refractive indices $u$ with almost complete transmission.

\section{Conclusion}\label{sec:conclusion}

We studied the transmission of light through quasiperiodic multilayers in dependency on the underlying structure, the incidence angle, and the light polarization. We obtained additional bands with almost complete transmission in the quasiperiodic systems at frequencies in the range of the photonic band gap of a periodic system for both types of light polarization. We found that with increasing incidence angle these bands bend towards higher frequencies and derived a relation in dependency on the metallic mean $\tau(a)$. This indicates that the bending is solely caused by the overall phase difference occurring during the passage of the stack. Further, in the quasiperiodic systems for p-polarized light the bands showed almost complete transmission near the Brewster's angle $\vartheta_\textrm{Br}$ in contrast to the results for s-polarized light. In quasiperiodic systems almost complete transmission occurs for a significant range of ratios $u$ in contrast to the periodic case.

\subsection*{Acknowledgements}

The authors thank the Stiftung der Deutschen Wirtschaft and EPSRC (grant EP/D058465) for funding the research.

\newcommand{\noopsort}[1]{} \newcommand{\printfirst}[2]{#1}
  \newcommand{\singleletter}[1]{#1} \newcommand{\switchargs}[2]{#2#1}

\end{document}